\documentclass[aip,jcp,reprint,amssymb,amsmath,superscriptaddress,groupedaddress,frontmatterverbose,]{revtex4-2}

\usepackage{amsmath}
\usepackage{amssymb}
\usepackage{bm}
\usepackage{color}
\usepackage{braket}
\usepackage{graphicx}
\usepackage{orcidlink}
\usepackage{epstopdf}
\usepackage{physics}
\usepackage{comment}
\usepackage{ulem}
\DeclareGraphicsExtensions{{.eps,.pdf,.png}}
\graphicspath{{./}}

\begin{document}
\title{Analysis of intramolecular modes of liquid water in two-dimensional spectroscopy: a classical hierarchical equations of motion approach}
\date{Last updated: \today}

\author{Ryotaro Hoshino\orcidlink{xxxxx-xxxx-xxxx-xxxx}}
\author{Yoshitaka Tanimura\orcidlink{0000-0002-7913-054X}}
\email[Author to whom correspondence should be addressed: ]{tanimura.yoshitaka.5w@kyoto-u.jp}
\affiliation{Department of Chemistry, Graduate School of Science,
Kyoto University, Kyoto 606-8502, Japan}

\begin{abstract}
Two-dimensional (2D) vibrational spectroscopy is a powerful means of investigating the structure and dynamics of complex molecules in condensed phases. However, even in theory, analysis of 2D spectra resulting from complex inter- and intramolecular motions using only molecular dynamics methods is not easy. This is because molecular motions comprise complex multiple modes, and peaks broaden and overlap owing to various relaxation processes and inhomogeneous broadening. On the basis of an anharmonic multimode Brownian oscillator model with nonlinear system--bath coupling, we have developed an approach that simulates 2D spectra, taking into account arbitrary modes of intermolecular and intramolecular vibrations simultaneously. Although only two-mode quantum calculations are feasible with this model, owing to high computational costs, here we restrict ourselves to the classical case and perform three-mode calculations. We demonstrate the applicability of our method by calculating 2D correlation infrared spectra of water for symmetric stretching, antisymmetric stretching, and bending modes. The quantum effects of these results are deduced by comparing 2D quantum spectra previously obtained for two intramolecular modes with those obtained using our classical approach under the same physical conditions. The results show that the 2D spectra calculated by separating the stretching modes into symmetric and asymmetric modes provide better descriptions of peak profiles, such as the splitting of cross-peaks.
\end{abstract}

\maketitle

\section{INTRODUCTION}
Interactions between solute molecules and solvent water have critical roles in many chemical and biological processes; for instance, high-frequency intramolecular modes facilitate bond formation and breakage of solute molecules, whereas low-frequency intermolecular modes contribute irreversible thermal excitation and relaxation.\cite{Ohmine_ChemRev93,OCSACR1999}  
Infrared (IR), THz, and Raman spectroscopies have been used to elucidate the spectroscopic line shapes of liquid water under various vibrational modes and interactions of these modes.
Specifically, IR has been used to study intramolecular OH stretching ($\sim$3600 cm$^{-1}$) and HOH intramolecular bending ($\sim$1600 cm$^{-1}$) motions, whereas THz and Raman approaches have been used for HB intermolecular translational (vibrational) motions (33--100 cm$^{-1}$) and HB intermolecular vibrational motions (33--400 cm$^{- 1}$).

Recent advances in two-dimensional (2D) experimental techniques, which imprint an additional time correlation on the system response, have expanded our ability to study both intermolecular and intramolecular modes in the 0--4000 cm$^{-1}$ frequency range. These techniques include: 2D IR spectroscopy,\cite{TokmakoffGeisslerforHB2003,ElsaesserDwaynePNAS2008,JansenPshenichnikov2009,Tokmakoff2016H2O,VothTokmakoff_St-BendJCP2017,Tokmakoff2022,Kuroda_BendPCCP2014,Cho2009,Hamm2011ConceptsAM} 2D THz--Raman spectroscopies,\cite{HammTHz2012,HSOT12JCP,Hamm2013PNAS,hamm2014,HammPerspH2O2017} and 2D THz--IR--visible light \cite{grechko2018,Bonn2DTZIFvis2021} (equivalent to 2D IR--IR--Raman) spectroscopy.\cite{IT16JCP}   These methods are critical for understanding the degree of coupling to surrounding molecules, often referred to as the ``bath''.\cite{TM93JCP,mukamel1999principles,TI09ACR,UT20JCTC} 

Molecular dynamics (MD) simulations\cite{YagasakiSaitoJCP20082DIR,JansenChoShinji2DVPerspe2019} and simulations using stochastic models with noise correlation functions evaluated from MD simulations have been used to analyze these 2D signals. \cite{SkinnerStochs2003,Skinner2005,JansenSkinnerJCP2010}
 However, no theoretical results have yet been obtained that satisfactorily explain the experimental results. This is likely to be because water interactions are complex; thus, intramolecular modes must be treated quantum mechanically,\cite{BowmanQM_QD2015,Paesan2018H2OCMD,Althorpe2019CMD,JianLiu2018H2OMP} 
 and any calculation of multidimensional spectra must properly account for the effects of quantum entanglement due to interactions with surrounding molecules.\cite{TS20JPSJ,KT04JCP,ST11JPCA}

In condensed phases, a molecular system interacts with its surrounding environment (bath) in a non-perturbative and non-Markovian manner.  Such interactions lead to quantum entanglement (bathentanglement)\cite{T20JCP}  between the solute molecule and the bath in the quantum case; this, in turn, has an effect on the linear response spectrum, in the form of changes in peak position and profile. The entanglement effect in nonlinear response spectra is particularly important; examples include photon echo in interactions with external laser fields that produce echo signals. In such cases, the calculation of 2D spectra requires the use of theoretical approaches such as stochastic theory\cite{SkinnerStochs2003,Skinner2005,JansenSkinnerJCP2010}, hierarchical equations of motion (HEOM), \cite{TS20JPSJ,KT04JCP,ST11JPCA,T20JCP,TT23JCP1,TT23JCP2} and 
the quasi-adiabatic path integral,\cite{Makri2D2010} which treat the thermal bath in a non-Markovian, non-perturbative, and non-factorized manner.

The only quantum theory that currently enables effective description of the intramolecular modes of water, including population relaxation, phase relaxation, and anharmonicity of modes with coupling between modes, is the multimode anharmonic system--bath model. This model was built on the basis of MD,\cite{TI09ACR,IIT15JCP,IT16JCP} specifically targeted to be solved with the HEOM.\cite{TS20JPSJ,KT04JCP,IT05JCP,IT07JPCA,IT06JCP,ST11JPCA} It can be used to accurately describe the effects of non-perturbative and non-Markovian dephasing and relaxation, as well as temperature effects relating to thermal equilibrium.\cite{T20JCP,T06JPSJ}
However, it is a computationally expensive approach, and, to date, it has only been successfully applied to consider two modes at a time.\cite{TT23JCP1,TT23JCP2} To properly understand energy and coherence transfer between intramolecular modes, at least three modes must be considered simultaneously: the OH antisymmetric and symmetric stretching modes and the HOH bending mode.

Although experimental techniques are rapidly advancing, the limitation of theoretical analysis to two modes remains a significant impediment to scientific progress. In this study, we perform simulations in the classical case, taking advantage of current graphics processing unit (GPU) technology to perform three-mode calculations. First, we compute 2D IR spectra of a liquid water model consisting of three primary intermolecular modes. 
Although it is not possible for us to directly incorporate quantum effects, we assess the influence of such effects by comparing the results of quantum and classical calculations for the two-mode scenario.

This paper is organized as follows.
In Section~\ref{sec:model}, we introduce the multimode anharmonic Brownian model and the classical HEOM in a Wigner space representation. The set of parameters used in the simulation is also given. In Section~\ref{sec:NumericalResults}, 2D IR correlation spectra are computed and analyzed for the two-mode and three-mode cases. Section~\ref{sec:conclusion} presents some concluding remarks.

\section{Theory}
\label{sec:model}
\subsection{Multimode anharmonic Brownian model for intramolecular modes}\label{sub:MMBO}
We consider a liquid water model consisting of one of the three primary intermolecular and intramolecular modes.  
These modes are described by dimensionless vibrational coordinates $\bm{q}=(q_1, q_{1'}, q_2)$. Each mode is independently coupled to the other optically inactive modes, which constitute a bath system represented by an ensemble of harmonic oscillators. The total Hamiltonian can then be expressed as\cite{IT16JCP,TT23JCP1,TT23JCP2} 
\begin{align}
\hat{H}_{tot}=  \sum_{s} \qty( \hat{H}_{A}^{(s)} +\hat{H}_{I}^{(s)} +\hat{H}_{B}^{(s)}) +  \sum_{s<s'} \hat{U}_{ss'}\qty(\hat{q}_s, \hat{q}_{s'}),
\label{sec:Total Hamiltonian}
\end{align}
where
\begin{align}
\hat{H}_{A}^{(s)}= \frac{\hat{p}_s^{2}}{2m_s} +\hat U_s(\hat{q}_s)
\label{sec:System Hamiltonian}
\end{align}
is the Hamiltonian for the $s$th mode, with mass $m_s$, coordinate ${\hat{q}_s}$, and momentum ${\hat p_s}$; and
\begin{align}
\hat U_s(\hat{q}_s)= \frac{1}{2} m_s \nu_s^2 \hat{q}_s^2 +\frac{1}{3!}g_{s^3}q_{s}^3
\label{sec: Potenential s}
\end{align}
is the anharmonic potential for the $s$th mode, described by the frequency  $\nu_s$ and cubic anharmonicity $g_{s^3}$. The mode--mode interaction between the $s$th and $s'$th modes is expressed as  
\begin{align}
\hat{U}_{ss'}(\hat{q}_s, \hat{q}_{s'}) = g_{s{s'}}\hat{q}_s\hat{q}_{s'} + \frac{1}{6}  \qty(g_{s^2s'}\hat{q}_s^2 \hat{q}_{s'} + g_{s{s'}^2} \hat{q}_s \hat{q}_{s'}^2 ),
\label{sec: Potential ss'}
\end{align}
where $g_{s{s'}}$ represents the second-order anharmonicity, and $g_{s^2s'}$ and $g_{s{s'}^2}$ represent the third-order anharmonicity.  The bath Hamiltonian for the $s$th mode is expressed as
\begin{align}
\hat{H}_{B}^{(s)}= \sum_{j_s}\qty[\frac{\hat{p}_{j_s}^{2}}{2m_{j_s}}+\frac{m_{j_s}\omega_{j_s}^{2}}{2}\qty(\hat{x}_{j_s}-\alpha_{j_s} \hat{V}_s(\hat{ q}_s) )^2],
\label{sec: Bath Hamiltonian}
\end{align}
where the momentum, coordinate, mass, and
frequency of the $j_s$th bath oscillator are given by ${p}_{j_s}$, ${x}_{j_{s}}$, $m_{j_{s}}$ and
$\omega _{{j_s}}$, respectively.  The system--bath interaction, defined as
\begin{align}
  {H}^{(s)}_{\mathrm{I}}&=- V_{s}({q_s})\sum _{j_s}\alpha _{j_s}{x}_{j_s},
  \label{eq:h_int}
\end{align}
consists of linear--linear (LL)
and square--linear (SL) system--bath interactions,
$V_{s}({q_s})\equiv V^{(s)}_{\mathrm{LL}}{q_s}+V^{(s)}_{\mathrm{SL}}{q_s}^{2}/2$, with coupling strengths $V^{(s)}_{\mathrm{LL}}$, $V^{(s)}_{\mathrm{SL}}$,
and $\alpha _{j_s}$.\cite{TS20JPSJ,KT04JCP,IT05JCP,IT07JPCA,IT06JCP,ST11JPCA,IIT15JCP} 
Whereas the LL interaction mainly contributes to energy relaxation, the LL+SL system--bath interaction causes vibration dephasing in the case of slow modulation, owing to frequency fluctuations in the system oscillations. \cite{TS20JPSJ,OT97PRE,T06JPSJ,TI09ACR}
The bath property is characterized by the spectral distribution function (SDF), defined as,
\begin{align}
J_s(\omega) \equiv \sum_{j_s} \frac{\alpha^2_{j_s}}{2 m_s \omega_{j_s}} \delta (\omega-\omega_{j_s}) .
\label{sec: SDF}
\end{align}
Here, we assume the Drude SDF expressed as
\begin{equation}
  J_s(\omega)=\frac{m_s \zeta_s}{2\pi}\frac{ \gamma_{s}^{2}\omega}{\omega^{2}+\gamma_s^{2}}, 
\label{eq:drude}
\end{equation}
where  $\zeta_s$ is the system--bath coupling strength, and  $\gamma_s$ represents
the width of the SDF for mode $s$, which relates to the
vibrational dephasing time, defined as  $\tau_s$ = ${1}/{\gamma_s}$.

The classical collective coordinate is written as $X_{s}$. The correlation function is then given by $\langle X_{s}(t)X_{s}(0)\rangle \propto e^{-\gamma_{s} \left|t\right|}$. This indicates that the bath oscillators interact with the system in the form of stochastic Gaussian noise with correlation time $t_0$, if the relaxation effect is ignored.\cite{TK89JPSJ1,T06JPSJ} 
This model has been used to derive predictions for 2D Raman,\cite{TS20JPSJ,T06JPSJ} 2D THz--IR\cite{IIT15JCP}, 2D IR--Raman,\cite{IT16JCP,TT23JCP1} and 2D IR \cite{IT06JCP,ST11JPCA,IT07JPCA,TI09ACR} spectra. 

In the past, various stochastic-theory-based models have been developed to analyze 2D IR spectra with respect to stretching modes of water; in these models, the noise amplitude $\Delta$ and correlation time $\tau_{s}$ were evaluated based on classical MD trajectories or quantum mechanics/molecular mechanics approaches. The stochastic model for a single stretching mode corresponds to the LL+SL anharmonic Brownian model for a single mode when the noise correlation is slow ($\gamma_{s} \ll \omega_0$) and in the high-temperature limit ($\beta\hbar\omega_0/2 \ll 1$) with respect to the mode frequency $\omega_0$.\cite{IT06JCP}  As the stochastic model ignores the effects of the population relaxation, the 0-1-0 and 0-1-2 peak profiles are more or less symmetric and not reproducible, for example, as in experimentally obtained 2D spectra in the mid-IR region.\cite{Tokmakoff2016H2O,VothTokmakoff_St-BendJCP2017,Tokmakoff2022} 

Incorporating the non-Condon effect\cite{SkinnerStochs2003,Skinner2005,JansenSkinnerJCP2010}
 or the effects of intermolecular hydrogen coupling among the surrounding molecules\cite{Skinner2005} within the framework of the stochastic model in the eigenstate representation results in the 0-1-0 and 0-1-2 peaks becoming asymmetric. However, the calculated 2D spectra continue to differ significantly from the observed 2D spectra in the mid-IR region.\cite{Tokmakoff2016H2O,VothTokmakoff_St-BendJCP2017,Tokmakoff2022}  
 By contrast, the LL+SL anharmonic Brownian model described in molecular coordinates facilitates investigation of vibrational relaxation and energy transfer under not only fluctuation but also dissipation at finite temperatures in both intramolecular and intermolecular modes, in a physically consistent manner.\cite{TS20JPSJ,T06JPSJ}  
Thus, the present model allows for more detailed analysis of 2D vibrational spectra compared with previous approaches.

For the two-mode case, we determined the parameter set for the present model to reproduce the 2D IR--Raman spectra obtained from the classical MD simulations and modified it to account for quantum effects.\cite{TT23JCP1,TT23JCP2}  As an example of a three-mode case, we consider here the following modes:
(1) OH antisymmetric stretching (anti-stretching), ($1'$) OH symmetric stretching (stretching), and (2) HOH bending (bending).  

Notably, both the stochastic model and the present model use parameters based on classical MD results. 
However, the parameters of the stochastic model were selected to reproduce solely the trajectory of the stretching motion,\cite{SkinnerStochs2003,Skinner2005,JansenSkinnerJCP2010}  whereas those of the present model were chosen to reproduce the entire profile of the 2D IR--Raman spectrum obtained from MD results.\cite{IT16JCP} Consequently, the present model encompasses various effects, including phase relaxation and population relaxation between modes. 
Although it is difficult to compare parameters between the two models, because they have been constructed in very different ways, the noise correlation functions for each mode are well defined and show relatively good agreement, although their amplitude is very different. 

\subsection{Classical hierarchal Fokker--Planck equations for a multimode system}
\label{sec:CHFPE}

To study the effects of thermal activation, relaxation, vibrational dephasing, the anharmonicity of modes, and the nonlinearities of the dipole moment in the 2D spectra within a unified framework, we required a kinetic equation that could treat thermal fluctuations as well
as dissipation in a non-perturbative, non-Markovian manner. 
For the LL+SL anharmonic Brownian model, the classical hierarchal Fokker--Planck equations (CHFPE) in the phase space for the system described by Eqs.~\eqref{sec:Total Hamiltonian}$-$\eqref{eq:drude}, developed for multidimensional vibrational spectra, can be expressed as\cite{T06JPSJ,IIT15JCP,IT16JCP}
\begin{align}
\label{eq:HEOM}
  \frac{ \partial{W^{(\bm{n})}(\bm{q}, \bm{p}; t)}}{\partial t}& = 
	 (\hat{L}(\bm{q}, \bm{p}) -\sum_{s} n_{s} \gamma_{s}) W^{(\bm{n})}(\bm{q}, \bm{p}; t) \nonumber \\
	& +\sum_{s}\hat{\Phi}_{s} W^{(\bm{n}+\bm{e}_{s})}(\bm{q}, \bm{p}; t)\nonumber \\
	&+\sum_{s}\hat{\Theta}_{s} W^{(\bm{n}-\bm{e}_{s})}(\bm{q}, \bm{p}; t),
\end{align}
where $W^{(\bm{n})}(\bm{q}, \bm{p}; t)$ is the Wigner distribution function (WDF).  As we are considering the case of three modes, the hierarchical elements are expressed here as $\bm{n}$ = $(n_1,n_2,n_3)$, where each $s$th mode element is denoted by a positive integer $n_s$, and $\bm{e}_{s}$ is the unit vector for the $s$th space.
 Note that $W^{(\bm{n})}(\bm{q}, \bm{p}; t)$ has physical meaning only when $\bm{n}$ = $(0,0,0)$; for other values of $\bm{n}$, it is an auxiliary WDF that indicates the non-perturbative, non-Markovian system--bath interactions.\cite{T06JPSJ,T20JCP} 
The classical Liouvillian $\hat{L}$ for the system Hamiltonian $H_{\rm sys}(\bm{q}, \bm{p}) \equiv \sum_{s} {H}_{A}^{(s)} + \sum_{s<s'} U_{ss'}\qty(\hat{q}_s, \hat{q}_{s'})$ can be expressed as
\begin{align}
\label{eq:CL_liouville}
\hat{L}(\bm{q}, \bm{p})  W(\bm{q}, \bm{p}) &\equiv \{ H_{\rm sys}(\bm{q}, \bm{p}) , W(\bm{q}, \bm{p})  \}_{\mathrm{PB}}, 
\end{align}
where $\{ \hspace{2mm},\hspace{2mm} \}_{\mathrm{PB}}$ is the Poisson bracket defined as
\begin{align}
\{ A, B \}_{\mathrm{PB}}  \equiv \sum_s \left( \frac{\partial A}{\partial q_s}\frac{\partial B}{\partial p_s} - \frac{\partial A}{\partial p_s}\frac{\partial B}{\partial q_s} \right)
\end{align}
for any functions $A$ and $B$.  Operators $\hat{\Phi}_{s}$ and $\hat{\Theta}_{s}$ represent the  energy exchange between the $s$th mode and the $s$th bath, respectively. They are expressed as\cite{TS20JPSJ,KT04JCP} 
\begin{align}
	\label{eq:Phi}
	\hat{\Phi}_{s}=\frac{ \partial V_s(q_s)}{\partial q_s} \frac{ \partial }{\partial p_s}
\end{align}
and
\begin{align}
	\label{eq:Theta}
	\hat{\Theta}_{s}=\frac{ m_s \zeta_s \gamma_s}{\beta}\frac{ \partial V_s(q_s)}{\partial q_s} \frac{ \partial }{\partial p_s}
        +\zeta_s \gamma_s p_s \frac{ \partial V_s(q_s)}{\partial q_s}, 
\end{align}
where $\zeta_s$ is the coupling strength, $\gamma_s$ is the inverse correlation time, and $T$ is the temperature.

The three-body response function of the dipole moment can be expressed as\cite{T06JPSJ} 
\begin{eqnarray}
\label{BO:2D}
R(t_3,t_2,t_1) = 
\iint d\bm{p}d\bm{q} {\mu}(\bm{q}) \mathcal{G}(t_3){ {\mu}(\bm{q})}^{\times}
\mathcal{G}(t_2) \nonumber \\
\times { {\mu}(\bm{q})}^{\times}\mathcal{G}(t_1){{\mu}(\bm{q})}^{\times}W^{\rm eq}(\bm{p},\bm{q}),
\end{eqnarray}
where the hyperoperator $^{\times}$ is defined as ${{\mu}(\bm{q})}^{\times}{W(\bm{p},\bm{q})} \equiv \{ {\mu}(\bm{q}), W(\bm{p},\bm{q}) \}_{\mathrm{PB}}$; $\mathcal{G}(t)$ is the Green's function for the time-evolution operator, as presented in Eq. \eqref{eq:HEOM}; and $W^{\rm eq}(\bm{p},\bm{q})$ is the equilibrium WDF expressed in terms of the hierarchical elements.\cite{T06JPSJ,IIT15JCP,IT16JCP} The dipole function is defined as
\begin{eqnarray}
\label{eq:dipole}
{\mu}(\bm{q})=\sum_{s} (\mu_s q_s +\mu_{ss} q_s^2)+ \sum_{s\ne s'} \mu_{s{s'}} q_s q_{s'}.
\end{eqnarray}
The non-Condon effects and mode-mixing effects of dipolar interactions are included in the model as $\mu_{ss} q_s^2$ and $\mu_{s{s'}} q_s q_{s'}$.

The 2D correlation IR spectrum is obtained by adding the two terms corresponding to the rephasing term $R_{NR}(t_3,t_2,t_1)$ and the non-rephasing term $R_{R}(t_3,t_2,t_1)$ 
with equal weights.\cite{2DCrrJonas2001,2DCrrGe2002,2DCrrTokmakoff2003}   A common definition of a 2D correlation spectrum can be expressed as\cite{2DCrrGe2002,2DCrrTokmakoff2003,TI09ACR} 
\begin{align}
&I_{\rm{C}}({\omega _3},{t_2},{\omega _1}) \nonumber \\ 
&= {\rm Im} \left\{ \int_0^\infty  \dd {t_1} \int_0^\infty  \dd {t_3}
 e^{ - i{\omega _1} {t_1} }  e^{ + i{\omega _3}{t_3}} 
 R_{\rm{R}} ({t_3},{t_2},{t_1})\right\}  \nonumber \\ 
&  + {\rm Im} \left\{ \int_0^\infty  \dd{t_1}  \int_0^\infty  \dd{t_3} {e^{i{\omega _1}{t_1}}}{e^{ + i{\omega _3}{t_3}}} R_{{\rm{NR}}}({t_3},{t_2},{t_1}) \right\}.\nonumber \\
  \label{2DCorr}
\end{align}
Although we could theoretically separate $R_{{\rm{R}}}$ and $R_{{\rm{NR}}}$ by choosing specific Liouville paths in the energy state model, this process is not easy in cases where the response functions must be calculated in phase space. Therefore, we eliminate the undesired rephasing contribution, using the Fourier transform of $t_2$ for $I_{\rm{C}}({\omega_3},{t_2},{\omega _1})$ with $R_{\rm{R}} ({t_3},{t_2},{t_1})=
 R_{{\rm{NR}}}({t_3},{t_2},{t_1})=R(t_3,t_2,t_1)$ to remove the oscillatory contribution in period $t_2$ with frequency 
 $2\nu_s$, where $\nu_s$ is the frequency of the target mode $s$.\cite{TI09ACR,HT08JCP,YagasakiSaitoJCP20082DIR,TT23JCP2}

\subsection{Integration of the CHFPE}

The CHFPE is time-integrated using the fourth-order Runge--Kutta method. For efficient parallel processing, the coordinate derivatives in the kinetic term are represented as a triple diagonal matrix. We then integrate Eq. \eqref{eq:HEOM} using the compact-finite-difference scheme\cite{LELE199216}, with the non-uniform mesh defined as follows:
\begin{align}
	\label{eq:nonuniform}
q=\frac {kq_{hl}} {(1.0+\beta \mathrm{e}^{  -  \frac{k^2}{\alpha^2}})}  (1.0+\beta \mathrm{e}^{    -   \frac{1.0}{\alpha^2}}),
\end{align}
where $\alpha$ and $\beta$ are parameters that characterize the non-uniform mesh, $k$ is an equidistant parameter in the range $-1.0 \le k \le1.0$, and $q_{hl}$  is the size of half the domain in the $q$ direction. For numerical integrations, the hierarchy is truncated to satisfy the condition $\delta_{tot}>\Delta_{\bm{n}}/N$, where $\delta_{tot}$ is the tolerance of the truncation, with $N=\sum_{s} n_s$ and 
\begin{align}
	\label{eq:Trunicate}
	\Delta_{\bm{n}}=\prod_s \frac{1}{{(n_s)}^{0.05}}  {(\frac{m_s \zeta_s}{\beta})}^{n_s}.
\end{align}
By adjusting the number of hierarchical elements, we can calculate the spectrum with the desired accuracy.

The time evolution in the CHFPE is processed using parallel cyclic reduction. The entire integration routine is coded in CUDA using the CUBLAS library and executed on a GPU without requiring memory transfer to the CPU. The source code that we developed was run on NVIDIA A100 (VRAM 80G) GPU boards hosted by PC with Intel XEON 6212U (24 cores) and took about 1 week to compute one 2D correlation spectrum on a single GPU. The VRAM used during calculations was 1~GB or less. The source code used in the present investigation will be provided in a forthcoming paper.

\section{Numerical results}
\label{sec:NumericalResults}

The method developed in this work can be used to simulate nonlinear vibrational spectra of intramolecular and intermolecular modes of any molecule via design of a multimode LL+SL Brownian model based on MD simulations and/or experimental results. Although the classical description is valid for intermolecular vibrational modes in which the thermal excitation at room temperature is close to the excitation frequency, intramolecular vibrational modes in which the vibrational excitation energy is much higher than the thermal excitation must be treated quantum mechanically. 
However, with current CPU power, only two-mode quantum calculations are possible.\cite{TT23JCP1,TT23JCP2} Therefore, in the present work, we limited ourselves to the classical case and used GPUs instead of CPUs to perform three-mode calculations. Quantum effects on the classical three-mode spectrum were then inferred by comparing the quantum two-mode spectra presented in refs. \onlinecite{TT23JCP1,TT23JCP2} with classical two-mode spectra obtained using the present software for the same model.

\subsection{The two-mode case: comparison with quantum results}

\begin{table*}[!tb]
 \caption{\label{tab:para1}Parameter values of the multimode intramolecular LL+SL BO model for (1) stretching and (2) bending modes obtained on the basis of refs. \onlinecite{TT23JCP1,TT23JCP2}. Here, we set the fundamental frequency to $\nu_{0}$ = 4000  cm$^{-1}$. The normalized parameters were defined as $\tilde{\zeta}_s \equiv (\omega_0/\omega_s)^2\zeta_s$, $\tilde{V}_{LL}^{(s)} \equiv (\nu_s/\nu_0)V_{LL}^{(s)}$, $\tilde{V}_{SL}^{(s)} \equiv V_{SL}^{(s)}$, $\tilde{g}_{s^3} \equiv (\nu_s/\nu_0)^3 g_{s^3}$,$\tilde{\mu}_{s} \equiv (\mu_0/\omega_s)\mu_{s}$, and $\tilde{\mu}_{ss} \equiv (\nu_0/\omega_s)^2 \mu_{ss}$. Anharmonic mode--mode coupling and dipole elements are given by $\tilde{g}_{1^21'}=0$, $\tilde{g}_{11'^2}=0.2$, and $\tilde{\mu}_{11'}=2.0 \times 10^{-3}$. }
\scalebox{0.9}{
\begin{tabular}{ccccccccc}
  \hline \hline
s&    $\nu_s$ (cm$^{-1}$) & $\gamma_s/\omega_0$ & $\tilde{\zeta}_s$ & $\tilde{V}_{LL}^{(s)}$ & $\tilde{V}_{SL}^{(s)}$ & $\tilde{g}_{s^3}$ & $\tilde{\mu}_{s}$  & $\tilde{\mu}_{ss}$\\
  \hline
1&   $3520$ & $5.0{\times}10^{-3}$ & $9$ & $ 0 $                     & $1.0$ & $-5.0{\times}10^{-1}$ & $ 3.3 $  & $1.2{\times}10^{-2}$\\
2&  $1710$ & $2{\times}10^{-2}$ & $0.8$ & $ 0 $                   & $1.0$ & $-7{\times}10^{-1}$ & $ 1.8 $ & $0$\\
  \hline
\end{tabular}
}
\end{table*}

\begin{figure}[htbp]
  \centering
  \includegraphics[keepaspectratio, scale=0.58]{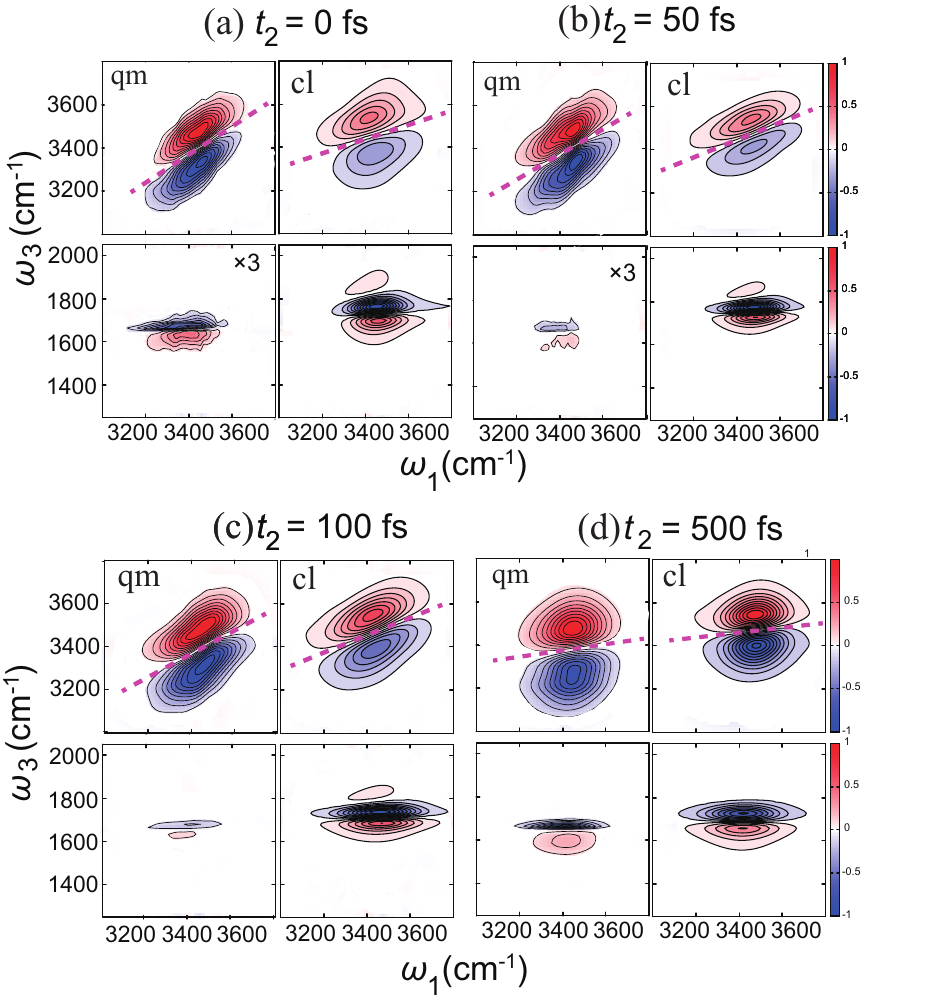}
  \caption{Third-order 2D correlation IR spectra for the two-mode case consisting of 
a stretching mode with $\nu_1$=3520~cm$^{-1}$ and a bending mode with $\nu_2$=1710 cm$^{-1}$. The upper panels show the stretching motion, whereas the lower panels show stretching$\rightarrow$bending motions for different $t_2$. All spectral intensities were normalized with respect to the absolute values of the maximum peak intensity of each diagonal peak. In each picture, the left panels show the quantum results from ref. \onlinecite{TT23JCP2}, and the right panels show the classical results. The direction of the nodal lines (red dashed lines) in the upper panel represents the extent of correlation between the vibrational coherences of the $t_1$ and $t_3$ periods. For clarity,  data for off-diagonal peaks were multiplied by 3 in the cases of $t_2=0$ fs and $50$ fs.
The quantum results were reproduced from H. Takahashi and Y. Tanimura, \href{https://doi.org/10.1063/5.0141181} {J. Chem. Phys.} {\bf 158}, 124108 (2023), with the permission of AIP Publishing.
}
  \label{fgr:2D2Mode}
\end{figure}

\begin{figure}[htbp]
  \centering
  \includegraphics[keepaspectratio, scale=0.56]{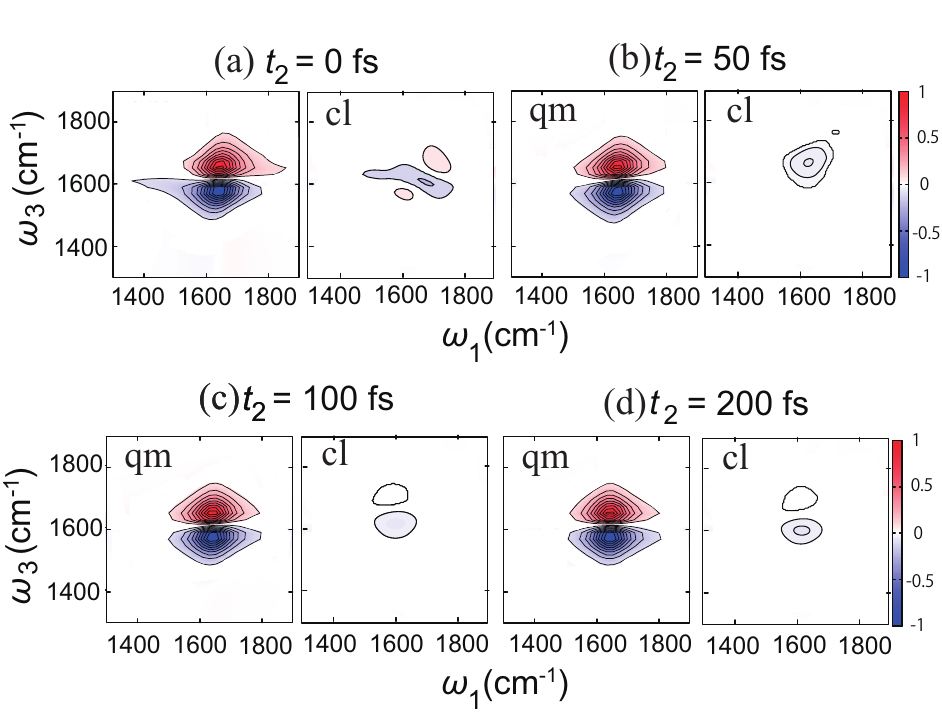}
  \caption{Third-order 2D correlation IR spectra of the bending mode calculated with stretching--bending coupling for different $t_2$. All spectral intensities were normalized with respect to the absolute maximum peak intensity in the $t_2=0$ case.
The left panels show the quantum results 
from ref. \onlinecite{TT23JCP2}, and the right panels show the classical results.
The quantum results were reproduced from H. Takahashi and Y. Tanimura, \href{https://doi.org/10.1063/5.0141181} {J. Chem. Phys.} {\bf 158}, 124108 (2023), with the permission of AIP Publishing.
}
  \label{fgr:2D2Modebend}
\end{figure}

First, we considered a two-mode case in which the stretching and anti-stretching modes were considered as a single mode with $\nu_1$=3520  cm$^{-1}$ that interacts with the bending mode with $\nu_2$=1710 cm$^{-1}$.\cite{IT16JCP} The parameter values of the simulations are given in Table \ref{tab:para1}.
 The two-mode model was developed to simulate a 2D IR--Raman spectrum using the CHFPE\cite{IT16JCP} and then modified to simulate 2D IR--Raman and 2D correlation IR spectra using quantum hierarchical Fokker--Planck equations.\cite{TT23JCP1, TT23JCP2}  Comparing previously obtained quantum mechanically calculated 2D IR spectra\cite{TT23JCP2} with those obtained here classically using the same model under the same conditions enabled us to identify the purely quantum effects in the 2D spectrum, thereby providing information for deducing quantum results from classical three-mode calculations.

Fig. \ref{fgr:2D2Mode} shows the 2D IR spectra for the stretching motion and the stretching$\rightarrow$bending motion obtained in the quantum and classical cases under the same physical conditions. Note that a similar comparison has previously been made for 2D 
IR--Raman spectroscopy.\cite{TT23JCP1} However, to investigate quantum effects in vibrational dephasing, it is necessary to perform the analysis for 2D IR. As in the 2D IR--Raman case, the discrepancy in peak positions between the classical and quantum scenarios could be attributed to the classical treatment of the anharmonic potential using a quantum mechanically constructed potential.\cite{CLQMdifference1981,BowmanQM_QD2015,HT11JPCB,ST11JPCA,CL2DIRBrumer2018,CLRepperts2023} 
In the classical case, the frequency is determined by the curvature at the bottom of the potential; in the quantum case, it is amplified by zero-point oscillations and can be determined by the difference between the ground state and the first excited state. 
The peak separation corresponding to the 0-1-0 and 0-1-2 transitions does not occur in classical simulations, because the energy is not discretized. 
However, the small peak separation between the (red) peak corresponding to the 0-1-0 transition and the (blue) peak corresponding to the 0-1-2 transition in the quantum results indicates that the anharmonicity of the modes was small, whereas the spectrum vanished without anharmonicity.\cite{ST11JPCA} 

The broadening of the peak profile in the $\omega_1$--$\omega_3$ diagonal direction resulted from the inhomogeneous distribution,\cite{TI09ACR,Hamm2011ConceptsAM} which is larger in the quantum case than in the classical case. This is because the intramolecular vibrational states in the classical case are localized at the bottom of the potential at room temperature, whereas in the quantum case, they spread out owing to the zero-point oscillation of the ground state. 

In the quantum case, the intensity of the stretching--bending cross-peak first decreased and then increased with time $t_2$, whereas in the classical case, the intensity decreased gradually with time. This was because in the quantum case, the cross-peak arises from the coherence between the stretching and bending modes, whereas in the classical case, it arises as a result of population relaxation from the stretching mode to the bending mode. As shown in Fig. \ref{fgr:2D2Modebend}, the intensity of the peak decreased monotonically.  By contrast, in the quantum case, the intensity did not change, and the nodal line of the peak exhibited a loss of coherence up to 50~fs.

Except for the region below $50$~fs, where the effects of quantum coherence became important, qualitative properties such as the phase relaxation time, which could be estimated from the positions of the nodes, did not differ significantly between the classical and quantum cases. This was because the anharmonicity of the intramolecular vibrational modes was small, such that no difference between the quantum and classical dynamical behaviors was apparent.

\subsection{The three-mode case: effect of the inter-stretch coupling}

\begin{table*}[!tb]
  \caption{\label{tab:para2} Parameter values of the multimode intramolecular LL+SL BO model for (1) asymmetric stretching, (1$'$) symmetric stretching, and (2) bending modes.
  Here, we set the fundamental frequency to $\omega_{0} = 4000$  cm$^{-1}$.}
\scalebox{0.9}{
\begin{tabular}{cccccccccc}
  \hline \hline
s&    $\nu_s$ (cm$^{-1}$) & $\gamma_s/\omega_0$ & $\tilde{\zeta}_s$ & $\tilde{V}_{LL}^{(s)}$ & $\tilde{V}_{SL}^{(s)}$ & $\tilde{g}_{s^3}$ & $\tilde{\mu}_{s}$  & $\tilde{\mu}_{ss}$\\
  \hline
1&   $3570$ & $5.0{\times}10^{-3}$ & $9$ & $ 0 $                     & $1.0$ & $-5.0{\times}10^{-1}$ & $ 3.3 $ & $1.2{\times}10^{-2}$\\
1'&   $3470$ & $5.0{\times}10^{-3}$ & $9$ & $ 0 $                     & $1.0$ & $-5.0{\times}10^{-1}$ & $ 3.3 $& $1.2{\times}10^{-2}$\\
2&  $1710$ & $2{\times}10^{-2}$ & $0.8$ & $ 0 $                   & $1.0$ & $-7{\times}10^{-1}$ & $ 1.8 $& $0$\\
  \hline
  \hline \hline\\
\end{tabular}
}
\end{table*}

\begin{table*}[!tb]
\caption{\label{tab:FitAll2}Parameter values for anharmonic mode--mode coupling and optical properties in the case of weak inter-stretch coupling.
}
\begin{tabular}{ccccccccccc}
  \hline \hline
  $\mathrm{s-s'}$ & $\tilde{g}_{ss'}$  & $\tilde{g}_{s^2s'}$ & $\tilde{g}_{s{s'}^2}$ & $\tilde{\mu}_{ss'}$ \\
  \hline
  $\mathrm{1-1'}$  & $-2.5{\times}10^{-3}$ & $0.16$ & $-2.1{\times}10^{-3}$ & $ 0$ \\
  $\mathrm{1-2}$  & $5{\times}10^{-8}$ & $-1.3{\times}10^{-2}$ & $-2{\times}10^{-4}$ & $2.0 \times 10^{-3}$ \\
  $\mathrm{1'-2}$  &  $-4{\times}10^{-4}$ &$6.2{\times}10^{-2}$ & $-6.0{\times}10^{-3}$ & $2.0 \times 10^{-3}$ \\
  \hline \hline\\
\end{tabular}
\end{table*}

\begin{table*}[!tb]
\caption{\label{tab:FitAll3}Parameter values for anharmonic mode--mode coupling and optical properties in the case of strong inter-stretch coupling.
Each mode--mode coupling variable was set to approximately twice that shown in Table \ref{tab:FitAll2}.
}
\begin{tabular}{ccccccccccc}
  \hline \hline
  $\mathrm{s-s'}$ & $\tilde{g}_{ss'}$  & $\tilde{g}_{s^2s'}$ & $\tilde{g}_{s{s'}^2}$ & $\tilde{\mu}_{ss'}$ \\
  \hline
  $\mathrm{1-1}'$  & $-5{\times}10^{-3}$ & $0.32$ & $-4.2{\times}10^{-3}$ & $ 0$ \\
  $\mathrm{1-2}$  & $10{\times}10^{-8}$ & $-2.6{\times}10^{-2}$ & $4{\times}10^{-4}$ & $2.0 \times 10^{-3}$ \\
  $\mathrm{1'-2}$  &  $-8{\times}10^{-4}$ &$1.2{\times}10^{-1}$ & $-1.2{\times}10^{-2}$ & $2.0 \times 10^{-3}$ \\
  \hline \hline\\
\end{tabular}
\end{table*}

\begin{figure}[htbp]
  \centering
  \includegraphics[keepaspectratio, scale=0.3]{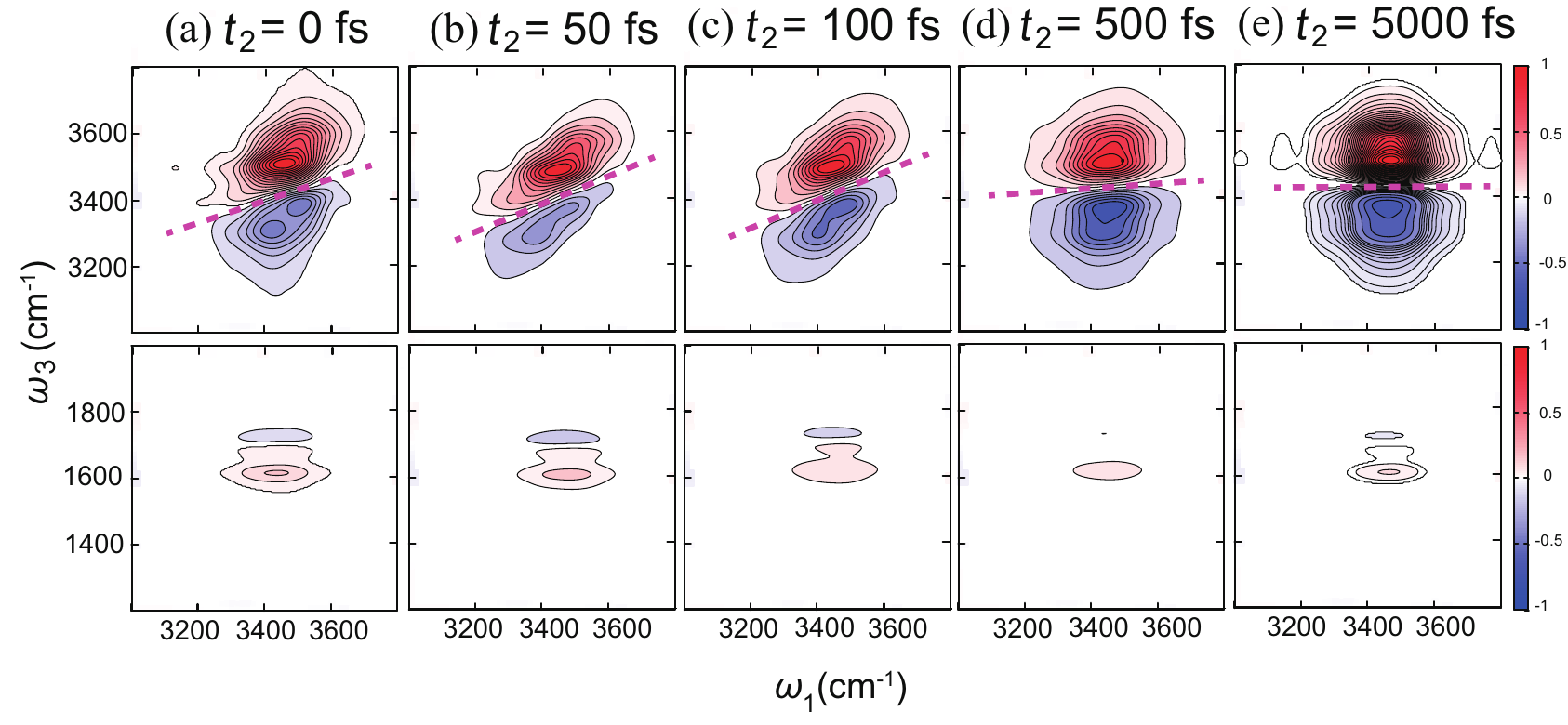}
  \caption{Third-order 2D correlation IR spectra for stretching and 
stretching$\rightarrow$bending motions calculated with the three-mode model consisting of (1) OH stretching with $\nu_1=3570$ cm$^{-1}$, ($1'$) OH anti-stretching with $\nu_{1'}=3470$ cm$^{-1}$, and (2) HOH bending with  $\nu_2= 1710$ cm$^{-1}$. The mode--mode coupling strength between these three modes was chosen to be weak (Table \ref{tab:FitAll2}).  The remaining parameters were the same as in Fig. \ref{fgr:2D2Mode}. 
All spectral intensities were normalized with respect to the absolute maximum peak intensity of each diagonal peak.
}
  \label{fgr:2D3Modelweak}
\end{figure}

\begin{figure}[htbp]
  \centering
  \includegraphics[keepaspectratio, scale=0.4]{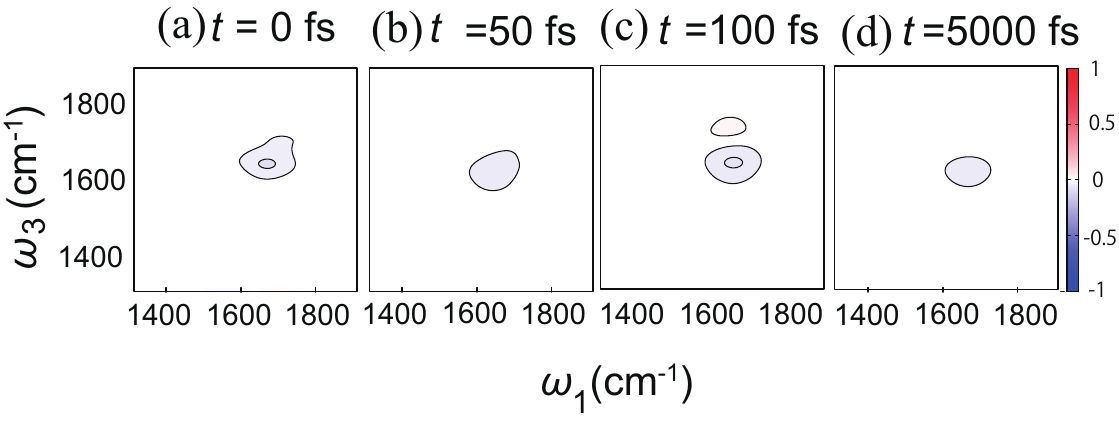}
  \caption{Results are shown for the same calculations as in Fig. \ref{fgr:2D2Modebend}, but for bending modes performed on the three-mode model for weak mode--mode coupling. As the peak intensity
  was weaker than that in Fig. \ref{fgr:2D3Modelweak}, the contour interval was tripled for emphasis.
}
  \label{fgr:2D3ModeBendWeak}
\end{figure}

\begin{figure}[htbp]
  \centering
  \includegraphics[keepaspectratio, scale=0.3]{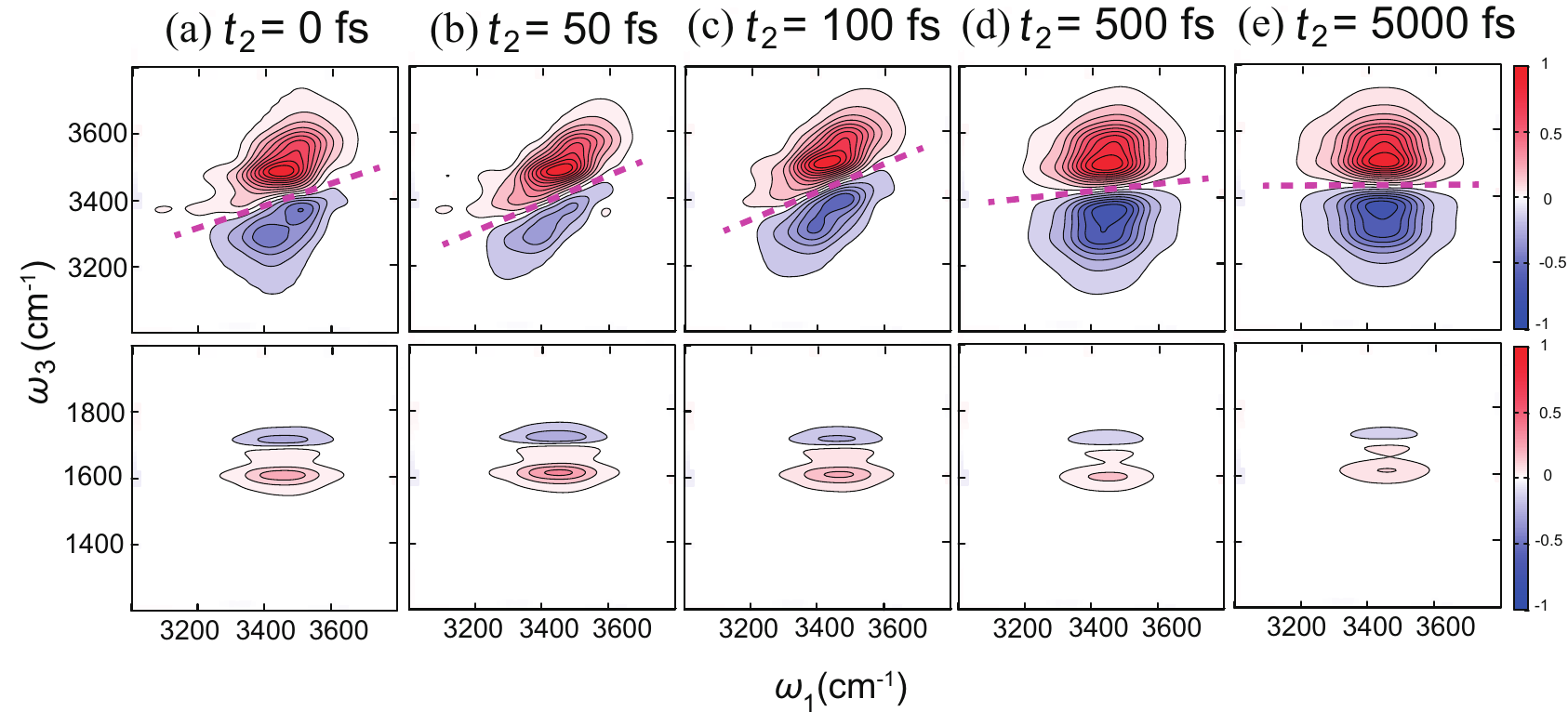}
  \caption{Results are shown for the same calculations as in Fig. \ref{fgr:2D3Modelweak}, except that we set the strong mode--mode coupling (Table \ref{tab:FitAll3}). All spectral intensities were normalized with respect to the absolute value of the maximum peak intensity of each diagonal peak.
}
  \label{fgr:2D3ModeStrong}
\end{figure}

\begin{figure}[htbp]
  \centering
  \includegraphics[keepaspectratio, scale=0.4]{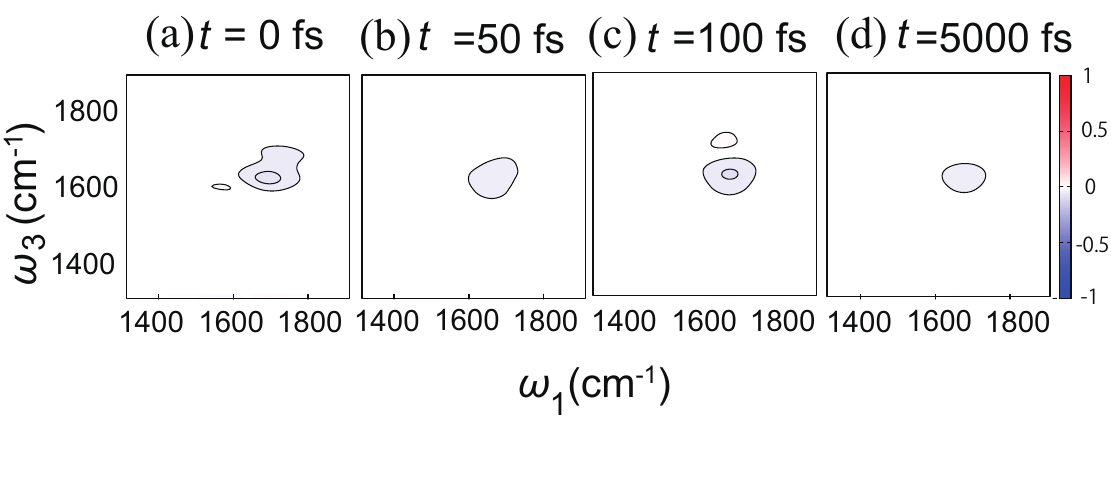}
  \caption{Results are shown for the same calculations as in Fig. \ref{fgr:2D2Modebend}, but for bending modes. As the peak intensity was weak compared with that in Fig. \ref{fgr:2D3ModeStrong}, the contour interval was tripled for emphasis.
} 
  \label{fgr:2D3ModeBendStrong}
\end{figure}

We next present the calculation results for the three-mode case, which include: (1) the OH stretching mode with $\nu_1$=3570 cm$^{-1}$, ($1'$) OH anti-stretching mode with $\nu_{1'}$=3470 cm$^{-1}$, and (2) HOH bending mode with  $\nu_2$=1710 cm$^{-1}$.  
As the stretching and anti-stretching modes could not be distinguished from the 2D spectral profiles, 
it was not feasible to ascertain the model parameters for each of the two modes and the coupling between them using the 2D IR--Raman-based method. Therefore, we determined these by referring to the parameters of a similar model for the Brownian SDF, which were obtained directly from the MD trajectories using a machine learning approach.\cite{UT20JCTC} However, given the inherent differences in models that include polarization, the parameters of two models differed significantly, including the coupling strength between the stretching--bending modes.
Therefore, we used only the ratio of the coupling strength between the three intramolecular modes obtained from the ML approach; its overall magnitude was used as a parameter, as given in Table \ref{tab:FitAll2} for a weak coupling case and in Table \ref{tab:FitAll3} for a strong coupling case. 
It is important to note that here the resulting coupling between the stretching modes was so strong that the relative strength of the coupling between the stretching and bending modes was weaker than in the two-mode case.

Fig. \ref{fgr:2D3Modelweak} shows 2D correlation IR spectra for the stretching modes and the stretching$\rightarrow$bending cross-peaks, whereas Fig. \ref{fgr:2D3ModeBendWeak} shows those for the bending mode in the case of weak mode--mode coupling (Table \ref{tab:FitAll2}). 
The remaining parameters used are given in Table \ref{tab:para2}. Here, we considered the two stretch modes separately;
the positive (red) 0-1-0 peaks of symmetric and antisymmetric stretching appear at 
frequencies of $( \omega_1, \omega_3)=(3570, 3570)$ and (3470, 3470) cm$^{-1}$, respectively, whereas the negative (blue) 0-1-2 peaks appear at $(3470, 3300)$ and (3570, 3400) cm$^{-1}$, respectively.
This was because in our classical simulations, each wave packet is localized at the bottom of the potential, so the peak was not broadened, whereas in the real system each peak was broadened and overlapped as a single peak owing to quantum effects.
 In addition, the inhomogeneous broadening was enhanced owing to the presence of two separate peaks compared with the two-mode case.  
 
As $t_2$ increased, the 2D peak profiles evolved in time from homogeneous to inhomogeneous distributions. The $t_2$ dependence of the stretching peaks was similar to 
that of the classical and quantum two-mode cases, with a vibrational dephasing time of about 500~fs. At $t_2 = 5000$~fs, we observed two diagonal peaks from the two stretching modes and two off-diagonal peaks from the transitions 
among them; thus, square-like plateaus appeared in the positive and negative peaks. 
To investigate the effects of the frequency difference between the two stretching modes, we reduced the difference by 20 cm$^{-1}$. The results, which are presented in Appendix \ref{sec:intera3480}, show that formation of square plateaus was suppressed.

The stretching$\rightarrow$bending cross-peaks are shown in the lower panels in Fig. \ref{fgr:2D3ModeBendWeak}. The intensities of these peaks were much weaker than those shown in Fig. \ref{fgr:2D2Mode}, because the anharmonic coupling between 1-2 and $1'$-2 was weaker than that in the two-mode case. As the stretching$\rightarrow$bending peaks arose from two paths, 0-1 and 0-$1'$, their peak profiles were elongated in the $\omega_1$ direction, whereas there was no elongation in the $\omega_3$ direction because the anharmonicities of the two stretching modes were chosen to be the same. As in the two-mode case, the intensity of the positive and negative peaks decreased monotonically with increasing $t_2$ in this classical calculation.

For $t_2$=100 and 5000~fs, we observed a small positive peak between the pronounced positive and negative peaks. To analyze this feature, we plotted the 2D IR of the bending mode (Fig. \ref{fgr:2D3ModeBendWeak}); we observed that the peak position of the bending mode in the $\omega_3$ direction was higher than those of the the stretching$\rightarrow$bending peaks. This indicated that the dominant contribution of the signal came from the 2-1 transition of the bending mode, which was lower than the 1-0 transition frequency owing to the anharmonicity of the bending mode. 
Given that $\mu_{1 {\rm} or 1'} \approx 2 \mu_2$, the second excited state of the bending mode was effectively excited by the stretching mode through anharmonic couplings $\tilde{g}_{2^2 1}$ and $\tilde{g}_{2^2 1'}$.
Consequently, the contribution of the 2-1 transition was greater than that of the 1-0 transition. The negative peak from the 2-3 transition and the positive peak from the 1-0 transition partially canceled each other out. The small peak between the positive and negative peaks in the stretching$\rightarrow$bending spectrum can be considered to be a remnant of the 1-0 transition peak.

The results for the strong mode--mode coupling case are shown in Fig. \ref{fgr:2D3ModeStrong}. Owing to the increased strengths of the 1-2 and $1'$-2 modes,  the stretching$\rightarrow$bending cross-peak became more pronounced compared with that shown in Fig. \ref{fgr:2D3Modelweak}, whereas the intensity of the stretching peaks was suppressed. Other than these changes, the overall peak profile remained largely unchanged as the mode--mode coupling strength increased. As the mode--mode coupling mechanism had a minimal impact on the excitation and de-excitation processes of the bending modes, the peak profiles shown in  Fig. \ref{fgr:2D3ModeBendStrong} remained consistent even with stronger coupling. These behaviors were also observed in the results shown in the Appendix \ref{sec:intera3480}, where the frequency difference between stretching and anti-stretching was reduced.

\section{Conclusion}
\label{sec:conclusion}
Although limited to the classical case, we have developed a CHFPE-based theory for an anharmonic multimode LL+SL Brownian model, enabling calculation of ultrafast 2D vibrational spectra by considering any three modes of inter- and intramolecular vibration. Using this framework, we have simulated 2D-correlated IR spectra for the stretching mode, the anti-stretching mode, and the bending mode of liquid water. Except for the region shorter than $50$~fs, where the effects of quantum coherence became important, the computed 2D spectra exhibited trends akin to those of experimental observations, including trends in qualitative properties such as increased inhomogeneous broadening and asymmetric profiles of the positive and negative peaks in the stretching modes.  However, we were unable to replicate the elongation of the negative stretching peak observed experimentally in the low-frequency direction.\cite{Tokmakoff2016H2O,VothTokmakoff_St-BendJCP2017,Tokmakoff2022} 
This elongation could be attributed to strong anharmonicity in the stretching mode potential, possibly a potential with a local minimum. The effects of combination bands of libration and bending modes in 2D IR\cite{Kuroda_BendPCCP2014} remain unexplored, as do the effects of anharmonicity and mode--mode coupling between low-frequency intermolecular modes in 2D THz--Raman spectra.\cite{HammTHz2012,HSOT12JCP,Hamm2013PNAS,hamm2014,HammPerspH2O2017} 

Our approach is also suitable for performing such calculations for any solute molecules in a solvent. Nevertheless, the selection of models and model parameters should be validated through comparison with results obtained using advanced experimental and simulation techniques.

\section*{Acknowledgments}
Y. T. thanks Peter Hamm, Andrei Tokmakoff, and Shinji Saito for stimulating discussions. 
Y.T. was supported by JSPS KAKENHI (grant no.~B21H01884).

\section*{Author declarations}
\subsection*{Conflict of Interest}
The authors have no conflicts to disclose.

\section*{Author Contributions}
{\bf Ryotaro Hoshino}: conceptualization (support); investigation (equal); software (lead); writing -- original draft (support); writing -- review and editing (support).
 {\bf Yoshitaka Tanimura}: conceptualization (lead); investigation (equal); software (support); Writing -- original draft (lead); writing – review and editing (lead).

\section*{Data availability}
The data that support the findings of this study are available from the corresponding author upon reasonable request.

\appendix

\section{Different frequency sets for stretching modes}
\label{sec:intera3480}

\begin{figure}[htbp]
  \centering
  \includegraphics[keepaspectratio, scale=0.3]{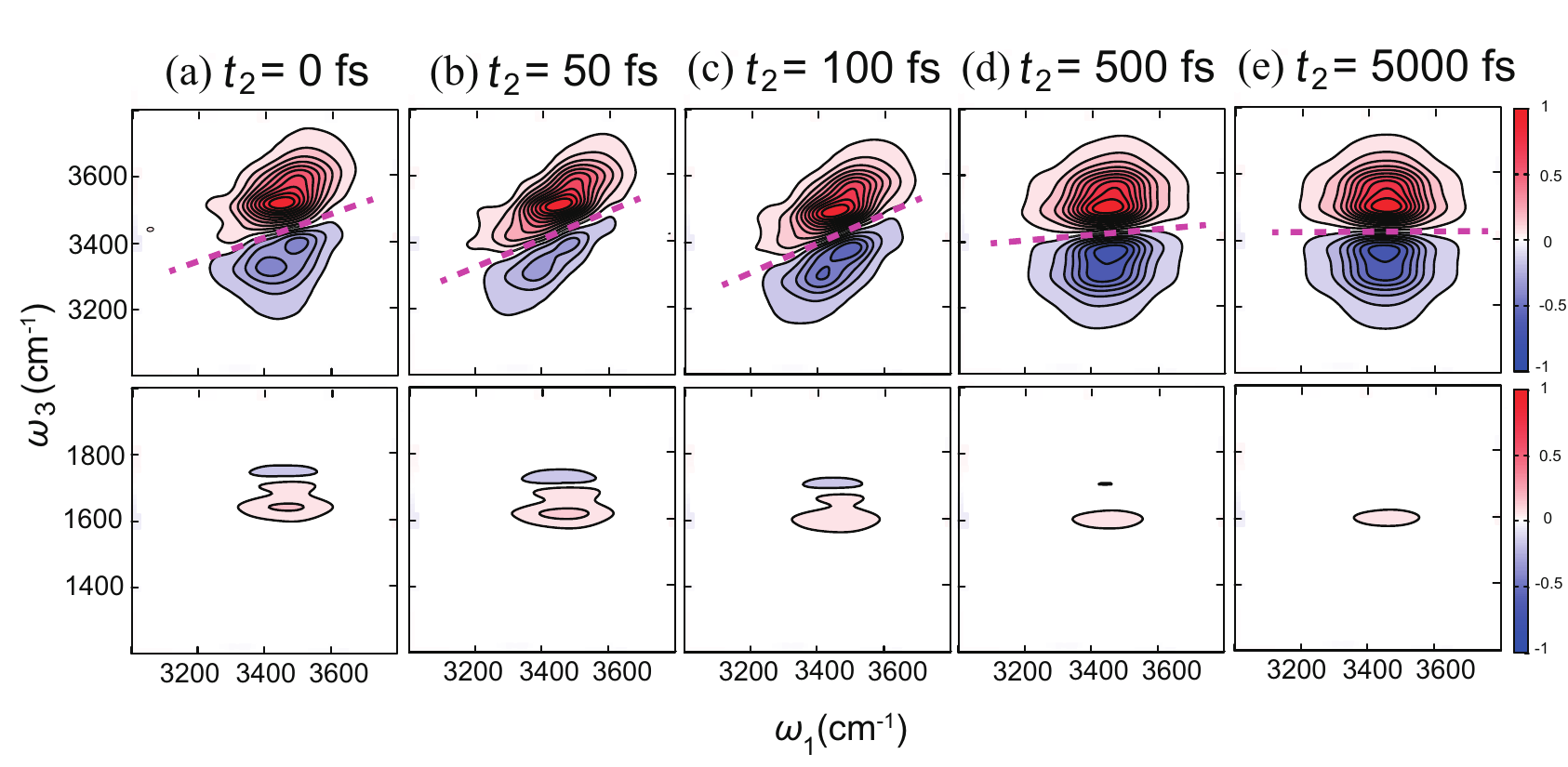}
  \caption{Third-order 2D correlation IR spectra for the three modes: (1) the OH stretching mode with $\nu_1=3560$ cm$^{-1}$, ($1'$) the OH anti-stretching mode with $\nu_{1'}$=3480 cm$^{-1}$, and (2) HOH bending motions with  $\nu_2$=1710 cm$^{-1}$. 
The interaction between the stretching and anti-stretching modes was weak (Table \ref{tab:FitAll2}).  The remaining parameters were the same as in Fig. \ref{fgr:2D2Mode}. 
All spectral intensities were normalized with respect to the absolute value of the maximum peak intensity of each diagonal peak.
}
  \label{fgr:2D3Model3560weak}
\end{figure}

\begin{figure}[htbp]
  \centering
  \includegraphics[keepaspectratio, scale=0.3]{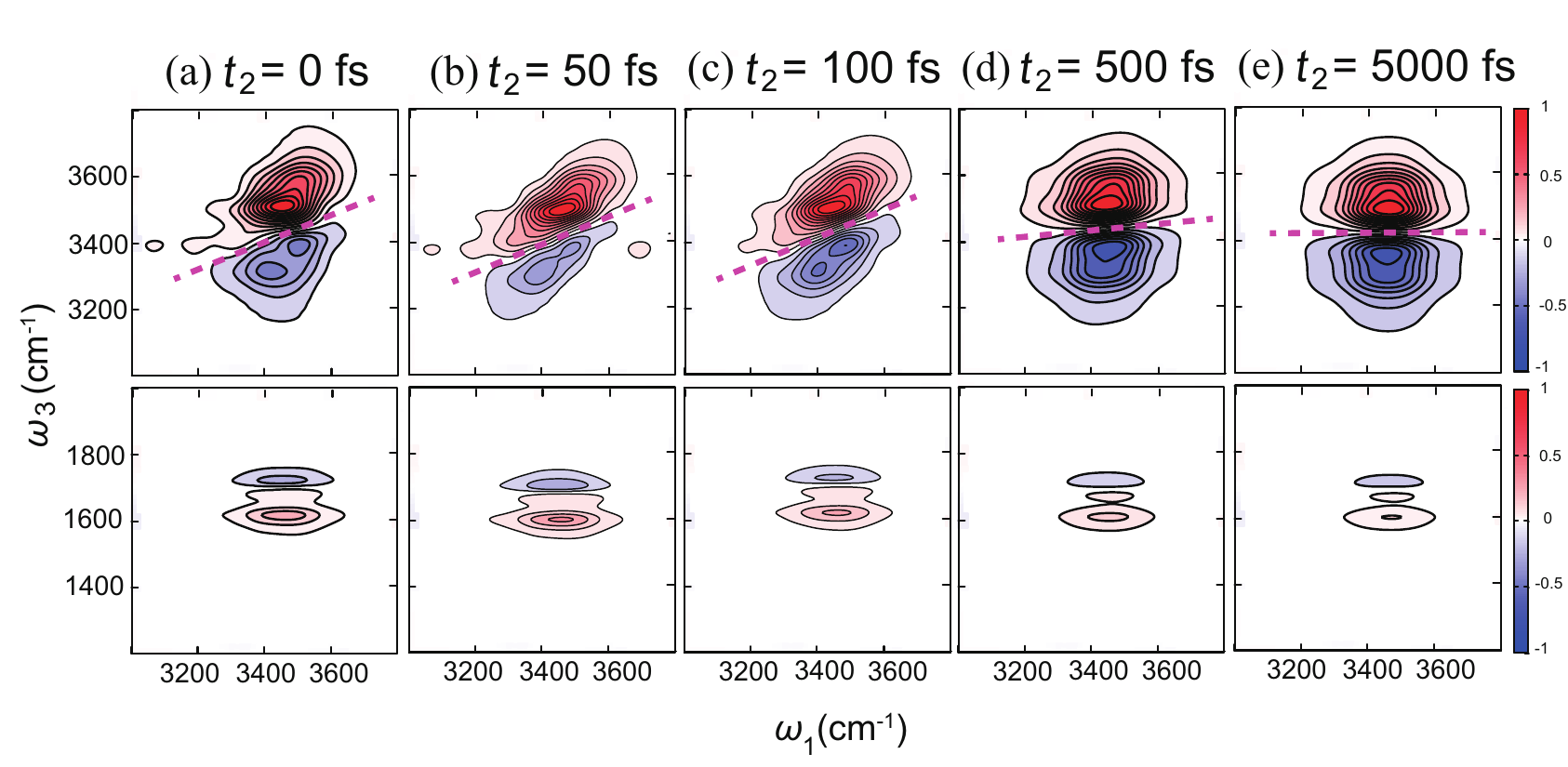}
  \caption{The same calculations were performed as in Fig. \ref{fgr:2D3Model3560weak}; besides this, we set the strong coupling (Table \ref{tab:FitAll3}). All spectral intensities were normalized with respect to the absolute value of the maximum peak intensity of each diagonal peak.
}
  \label{fgr:2D3Mode3560Strong}
\end{figure}

In addition to the coupling strength between the two stretching modes, we modified their fundamental frequencies (Figs. \ref{fgr:2D3Modelweak} and \ref{fgr:2D3ModeStrong}). Specifically, (1) the OH stretching mode was changed from $\nu_1$=3570 to 3560 cm$^{-1}$, and ($1'$) the anti-stretching mode was altered from $\nu_{1'}$ =3470 to 3480  cm$^{-1}$. All other parameters remained consistent with those shown in Figs. \ref{fgr:2D3Modelweak} and \ref{fgr:2D3ModeStrong}. 

The behaviors of the 2D peak profiles shown in Figs. \ref{fgr:2D3Model3560weak} and \ref{fgr:2D3Mode3560Strong} did not differ significantly from those shown in Figs. \ref{fgr:2D3Modelweak} and \ref{fgr:2D3ModeStrong}. However, the vibrational coherence, indicated by the nodal line at $t_0=0$, slightly increased as the resonance frequency decreased, because the two resonance peaks were closer together. They merged at a smaller $t_2$ owing to vibrational dephasing. The formation of a square-like plateau in the positive and negative peaks at $t_2$= 5000~fs was also suppressed.

For the stretching$\rightarrow$bending spectra, the three peaks could be more clearly observed because of the more effective energy transfer from the two stretching modes to the second excited state of the bending mode.

\bibliography{tanimura_publist,HT24,TT23}

\end{document}